\documentclass[aps,prl,reprint,superscriptaddress]{revtex4-1}
\usepackage{graphicx,graphics,amsfonts,amsmath,amssymb}

\usepackage{color}
\usepackage{verbatim}
\usepackage{tikz}

\begin{document}

%\preprint{}


\title{Coherent Microwave Control of Ultracold $^{23}$Na$^{40}$K Molecules}

\author{Sebastian A.~Will}
\affiliation{MIT-Harvard Center for Ultracold Atoms, Research Laboratory of Electronics, and Department of Physics, Massachusetts Institute of Technology,
Cambridge, Massachusetts 02139, USA }
\author{Jee Woo Park}
\affiliation{MIT-Harvard Center for Ultracold Atoms, Research Laboratory of Electronics, and Department of Physics, Massachusetts Institute of Technology,
Cambridge, Massachusetts 02139, USA }
\author{Zoe Z.~Yan}
\affiliation{MIT-Harvard Center for Ultracold Atoms, Research Laboratory of Electronics, and Department of Physics, Massachusetts Institute of Technology,
Cambridge, Massachusetts 02139, USA }
\author{Huanqian Loh}
\affiliation{MIT-Harvard Center for Ultracold Atoms, Research Laboratory of Electronics, and Department of Physics, Massachusetts Institute of Technology,
Cambridge, Massachusetts 02139, USA }
\affiliation{Center for Quantum Technologies, National University of Singapore, 3 Science Drive 2, Singapore 117543}
\author{Martin W.~Zwierlein}
\affiliation{MIT-Harvard Center for Ultracold Atoms, Research Laboratory of Electronics, and Department of Physics, Massachusetts Institute of Technology,
Cambridge, Massachusetts 02139, USA }
 
\date{\today}

%%%%%% ABSTRACT

\begin{abstract}
We demonstrate coherent microwave control of rotational and hyperfine states of trapped, ultracold, and chemically stable $^{23}$Na$^{40}$K molecules. Starting with all molecules in the absolute rovibrational and hyperfine ground state, we study rotational transitions in combined magnetic and electric fields and explain the rich hyperfine structure. Following the transfer of the entire molecular ensemble into a single hyperfine level of the first rotationally excited state, $J{=}1$, we observe collisional lifetimes of more than $3\, \rm s$, comparable to those in the rovibrational ground state, $J{=}0$. Long-lived ensembles and full quantum state control are prerequisites for the use of ultracold molecules in quantum simulation, precision measurements and quantum information processing.
\end{abstract}

\maketitle

%%%%%% INTRO

Ultracold molecules with large electric dipole moments hold great promise as a novel platform for quantum state-resolved chemistry~\cite{Quemener:2012}, precision measurements of fundamental constants~\cite{ACME2014, demille2008enhanced, zelevinsky2008precision}, quantum computation~\cite{demi02quantum} and quantum simulation \cite{Carr09mol,Baranov2012}, as well as for the realization of new states of dipolar quantum matter~\cite{krem09coldmolecules,Carr09mol,Baranov2012}. Essentially all anticipated applications depend on the ability to coherently control the quantum state of molecules, implying full control over electronic, vibrational, rotational and nuclear spin degrees of freedom~\cite{Quemener:2012}. With the recent production of dipolar molecules at sub-microkelvin temperatures~\cite{ni08polar, Takekoshi2014RbCs,Molony2014,Park2015:2, GuoNaRb2016}, this full quantum control has come into experimental reach for an entire ensemble of trapped molecules~\cite{Ospelkaus2010}.

Controlling the rotational states of molecules is directly linked to the control over long-range dipolar interactions~\cite{Buchler2007, Micheli:2007,gorshkov2008, avdee2006, avdee2012, quemener2016,Yan:2013}. Indeed, no state of definite parity can possess a dipole moment, but creating a superposition of opposite-parity rotational states induces one. Such a superposition can be achieved either via applying electric fields, or by coherently driving a microwave transition between rotational states. The potential applications for such coherent control range from quantum simulation of spin Hamiltonians~\cite{Barnett2006, micheli2006toolbox, Gorshkov:2011:2} to the realization of topological superfluidity~\cite{Cooper:2009}. Also, interaction control is expected to facilitate direct evaporative cooling of ultracold molecules~\cite{gorshkov2008, avdee2012, wang2015}.

In particular, for quantum information applications and many-body physics with dipolar molecules, a long lifetime of molecules in their individual quantum states is a key requirement. This is a prerequisite for having a large number of possible gate operations and for equilibration into novel phases, respectively. For ultracold chemically reactive molecules, losses can be prevented by isolating molecules in individual wells of an optical lattice~\cite{Chotia:2012}. Long lifetimes of several seconds in a bulk trapped sample of ultracold, rovibrational ground state molecules have been demonstrated for chemically stable, fermionic $^{23}$Na$^{40}$K~\cite{Park2015:2}. Whether a collection of trapped ultracold molecules in a rotationally excited state could have similarly long collisional lifetimes was thus far unknown.

%% SUMMARY

In this Letter, we demonstrate coherent microwave control over the rotational and hyperfine states of  ultracold $^{23}$Na$^{40}$K molecules. The experiment begins with the preparation of a spin-polarized  ensemble of fermionic $^{23}$Na$^{40}$K molecules in the absolute ground state~\cite{Park:2012, wu2012NaK, Park2015:1, Park2015:2}. First, we create weakly bound Feshbach molecules from an ultracold Bose-Fermi mixture of $^{23}$Na and $^{40}$K atoms in the vicinity of a Feshbach resonance. Subsequently, the Feshbach molecules are transferred to the rovibrational ground state via stimulated Raman adiabatic passage (STIRAP), coherently bridging an energy difference of $k_{\rm B} \times 7500\, \rm K$. The powers, frequencies and polarizations of the Raman lasers are optimized for efficient coupling into the lowest energy hyperfine state~\cite{Park2015:2}. The ultracold ensemble contains about $2\,\times 10^3$ molecules, all in the same internal quantum state, trapped in an optical dipole trap at a typical peak density of $1.1 \times 10^{11}\, \rm cm^{-3}$ and temperature of $400\, \rm nK$. For detection, we apply STIRAP in reverse, transferring the molecules back to the Feshbach state, where an absorption image is taken using light resonant with the atomic cycling transition of $^{40}$K \cite{wu2012NaK}.

%%%%%% SPECTRA VS. B FIELD

%%%%%%% FIGURE 1 - Cartoon and spectrum %%%%%%%
\begin{figure}
  \begin{center}
  \includegraphics[width=1\columnwidth]{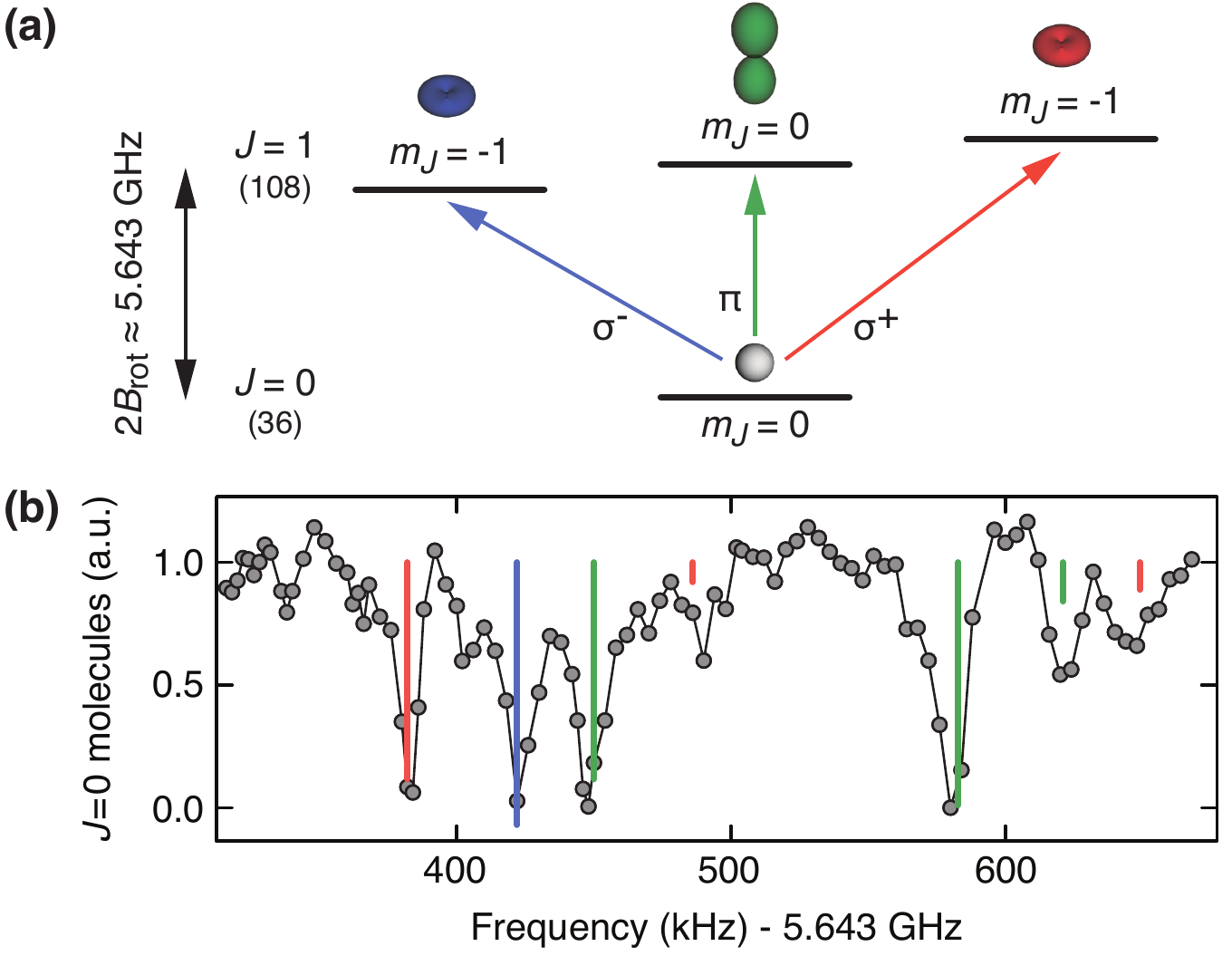}
  \end{center}
  \caption{\label{fig:cartoon}
(color online). Rotational excitations of $^{23}$Na$^{40}$K molecules from the rovibrational ground state. The rotational ground state $J{=}0$ is coupled to the excited $J{=}1$ states using microwave radiation with $\pi$, $\sigma^\pm$ polarization. (a) Simplified schematic omitting hyperfine interactions, such that $m_J$ is a good quantum number. In the full description, hyperfine interaction mixes $m_J$ with the nuclear spin states, resulting in a total of 108 hyperfine levels in $J=1$. (b) Spectrum of rotational transitions between the lowest hyperfine state of $J{=}0$ and $J{=}1$ at $B=216.6$ G. The position (height) of each vertical bar indicates the transition frequency (strength) according to our theoretical model. The calculated $\pi$-transition strength is scaled up by 1.9, accounting for the radiation characteristics of our antenna. During microwave exposure, the optical dipole trap is switched off to avoid differential Stark shifts between ground and excited rotational states. 
}
\end{figure}
%%%%%%%%%%%%%%%%%%%%%%%%%%%%%%%%

%%%%%%% FIGURE 2 - Spectra vs. B field %%%%%%%
\begin{figure}
  \begin{center}
  \includegraphics[width=1 \columnwidth]{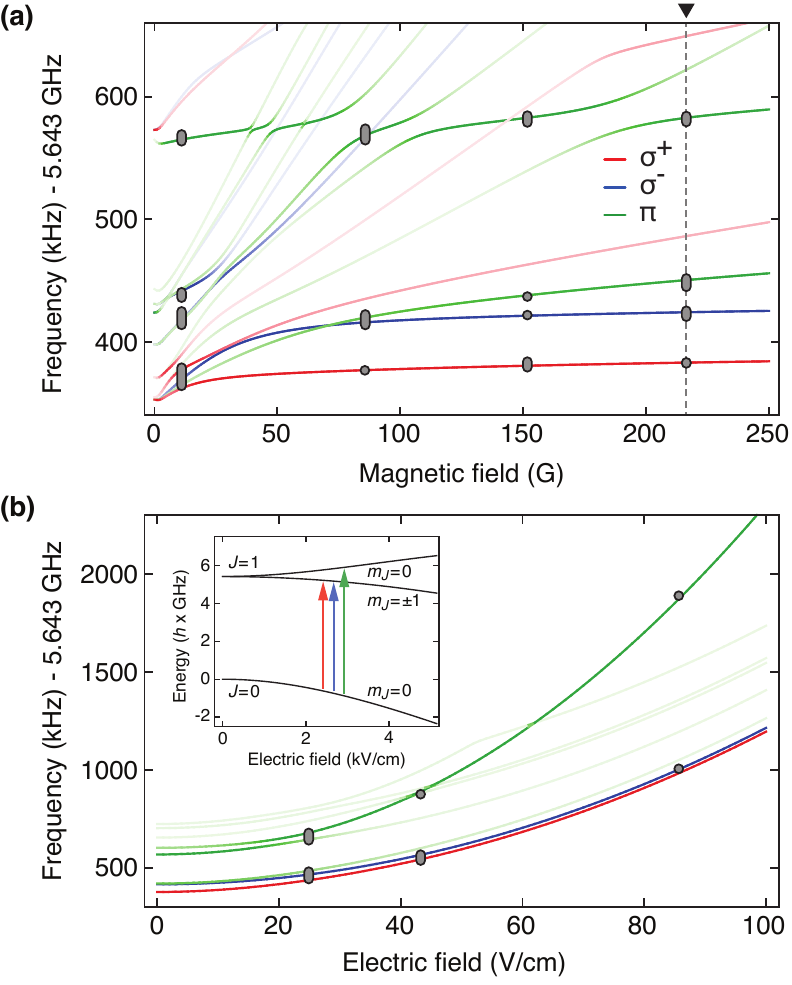}
  \end{center}
  \caption{\label{fig:Bfield} (color online). Microwave spectroscopy on $^{23}$Na$^{40}$K ground state molecules in magnetic and electric fields. (a) Gray markers indicate the observed transition frequencies at magnetic fields of 11.0 G, 85.6 G, 151.8 G, and 216.6 G. The vertical extent of the markers corresponds to the full width at half maximum of the observed resonances. Colored lines show the calculated transition frequencies for $\sigma^{-}$ (blue), $\pi$ (green), and $\sigma^{+}$ (red) transitions using best-fit parameters. The vertical dashed line indicates the spectrum of Fig.~\ref{fig:cartoon}(b). (b) Gray markers show the observed transition frequencies at electric fields of 24.8 V/cm, 43.2 V/cm, and 87.4 V/cm, simultaneously applied with a magnetic field of 85.6 G in the same direction. The observed Stark shift is used to calibrate the electric field assuming an electric dipole moment of $d=2.72$ D \cite{Gerdes2011}. The inset shows the expected Stark shift of $J{=}0$ and $J{=}1$ states for electric fields up to 5 kV/cm.
}
\end{figure}
%%%%%%%%%%%%%%%%%%%%%%%%%%%%%%%%

%% HYPERFINE STRUCTURE IN SINGLET 

The absolute ground state of bialkali molecules is an electronic spin-singlet state, ${\rm X}^1\Sigma^+$. Therefore, the hyperfine structure arises solely from the nuclear spins of $^{23}$Na and $^{40}$K, $I_{\rm Na} {=} 3/2$ and $I_{\rm K} {=} 4$, and their interplay with the rotation of the molecule. This leads to $(2I_{\rm Na}{+}1)(2I_{\rm K}{+}1) = 36$ hyperfine states in the rotational ground state $J{=}0$, and $(2J{+}1)(2I_{\rm Na}{+}1)(2I_{\rm K}{+}1) = 108$ states in the first rotationally excited state $J{=}1$. Already for magnetic fields above $ 2\,\rm G$, the hyperfine structure of the rotational ground state $J{=}0$ is dominated by the Zeeman effect of the $^{23}$Na and $^{40}$K nuclei \cite{Park2015:1}; in this regime, the states $  {\rm X}^{1}\Sigma^+ | v{=}0, J{=}0, m_{J}{=}0, m_{I_{ \rm Na}}, m_{I_{ \rm K}}\rangle$ form a good basis, with $v$ being the vibrational quantum number, and $m_J$,  $m_{I_{ \rm Na}}$, $m_{I_{ \rm K}}$ the quantum numbers associated with the component of $\vec{J}$, $\vec{I}_{\rm Na}$ and $\vec{I}_{\rm K}$ along the direction of the magnetic field. The lowest energy hyperfine state is $ {\rm X}^{1}\Sigma^+ | v{=}0, J{=}0, m_{J}{=}0, m_{I_{ \rm Na}}{=}3/2, m_{I_{ \rm K}}{=}{-}4\rangle$, abbreviated by $|0 , 0, 3/2, -4\rangle$ in the following. A spin-polarized ensemble of $^{23}$Na$^{40}$K molecules in the rovibrational ground state is stable against two-body chemical reactions \cite{zuch10mol}, and inelastic losses are suppressed as a consequence of fermionic quantum statistics. 

%% ROTATIONAL TRANSITIONS:

The frequencies of rotational transitions from $J{=}0$ to $J{=}1$ are about $2 B_{\rm rot} {\approx}5.643\, \rm GHz$, where $B_{\rm rot}$ is the rotational constant of $v{=}0$. If we were to neglect nuclear spins, the $J{=}1$ state would split into three sub-levels with $m_{J}{=} 0, \pm1$, giving rise to three electric dipole-allowed transitions from $J{=}0$ to $J{=}1$, as schematically shown in Fig.~\ref{fig:cartoon}(a). However, hyperfine interactions strongly couple rotation and nuclear spin, and additional states in the $J{=}1$ manifold with different nuclear spin projections become accessible. The only quantum number that remains good at all magnetic and electric fields is the projection of the total angular momentum $m_F = m_{J}{+}m_{I_{\rm Na}}{+}m_{I_{\rm K }}$. From the initial state, only those $J{=}1$ hyperfine states that satisfy the selection rule $\Delta m_{F}{=} 0, \pm1$ can be reached using $\pi, \sigma^{\pm}$ polarization, respectively.

%%%%%% TABLE 1 - Molecular constants

\begin{table}%[ht]
\centering
\begin{tabular}{l | l | l}
  Constant & Value & Reference \\
  \hline\hline
  $g_{\rm Na}$ & 1.477 & \cite{arim77}\\
  $g_{\rm K}$ & $-0.324$ & \cite{arim77}\\
  $g_{\rm rot}  $ &  0.0253(2) & \cite{Brooks:1972}\\
  $B_{\rm rot}$ (GHz)  & 2.821735 & \cite{Russier2000NaK}\\
  & 2.8217297(10) & This work \\
  $d$ (Debye) &  2.72(6) &\cite{Gerdes2011}\\
  $(eqQ)_{\rm Na}$ (MHz) & $-0.134(8)$& \cite{Brooks:1972}\\ %for 23Na39K (Ramsey's paper)
  & $-0.171(3)$& \cite{Dagdigian1972}\\ %for 23Na39K
  & $-0.187(35) $ & This work \\
  $(eqQ)_{\rm K}$ (MHz) & $0.893(3)$&\cite{Dagdigian1972,Jones1972}\\ %for 23Na39K; Jones paper has scaling factor: 718(2) * 1.244(2) 
  & $0.899(20)$ & This work \\
  $c_{4}  {\rm (Hz)}$ &  $-466.2$& \cite{Aldegunde:2015}\\
  & $-409(10)$ & This work\\
  \hline
\end{tabular}
\caption{\label{table:const} Constants of the molecular Hamiltonian for $^{23}$Na$^{40}$K. An improved value for $B_{\rm rot}$, as well as the nuclear quadrupole constants $(eqQ)_{\rm{Na}}$ and $(eqQ)_{\rm{K}}$ are obtained from a least squares fit to all resonances of Fig.~\ref{fig:Bfield}(a) that can be uniquely assigned to a single transition. The respective standard errors are determined using a bootstrap method, involving the resampling of residuals. The rotational $g$-factor $g_{\rm rot}$ and the scalar nuclear spin-spin constant $c_4$ are not varied in the fit; their impact is negligible in the investigated parameter regime.}
\end{table}
%%%%%%%%%%%%%%%%%%%%%%%%%%%%%%%%

%% OBSERVED SPECTRA AND MODEL

We drive rotational transitions on the trapped, spin-polarized ensemble and perform microwave spectroscopy to resolve the hyperfine structure of the first excited rotational state for magnetic fields up to $220 \,\rm G$ and electric fields up to $90\,\rm V/cm$. A typical spectrum is displayed in Fig.~\ref{fig:cartoon}(b), obtained by monitoring the remaining population in the absolute ground state $|0,0, 3/2,-4\rangle$ after microwave exposure, showing four dominant resonances. 

Figure~\ref{fig:Bfield}(a) summarizes the observed microwave transitions as a function of magnetic field, while the electric field is zero. The spectra are well described by a theoretical model of the hyperfine interaction in ${\rm X}^1\Sigma^+$, given by the molecular Hamiltonian $H_{\rm{mol}}{=}H_{\rm{rot}}{+}H_{\rm{Z}}{+}H_{\rm{hf}}$ \cite{Brown:2003,Aldegunde2008}. Here, $H_{\rm{rot}} = B_{\rm rot} h J (J+1)$ is the rotational contribution. $H_{\rm{Z}} = -\mu_{\rm N} (g_{\rm rot} m_J + g_{\rm Na} m_{I_{\rm Na}} + g_{\rm K} m_{I_{\rm K}}) B$ captures the Zeeman effect caused by the nuclear magnetic moments of $^{23}$Na and $^{40}$K and the much weaker rotational magnetic moment $g_{\rm rot} \mu_{\rm N}$, with $\mu_{\rm N}$ being the nuclear magneton and $B$ the magnetic field strength. 
While $H_{\rm rot}$ and $H_{\rm Z}$ are diagonal in the uncoupled basis, the hyperfine interactions are not. The two relevant hyperfine contributions are \mbox{$H_{\rm{hf}} = - \sum_{i= {\rm Na,\,K}} e ({\bf \nabla E})_i \cdot {\bf Q}_i + c_4 \vec{ I}_{\rm Na}\cdot \vec{ I}_{\rm K} $.} The first term vanishes for $J{=}0$, but for $J{=}1$ it is the dominant interaction. It describes the interaction of the intramolecular electric field gradient $({\bf \nabla E})_i$ at nucleus $i$ with the respective nuclear electric quadrupole moment $e {\bf Q}_i $, where $e$ is the electron charge. Matrix elements of $- e ({\bf \nabla E})_i \cdot {\bf Q}_i$ are proportional to the quadrupole coupling constant $(eqQ)_i$. The second term denotes the relatively weak scalar nuclear spin-spin interaction,  present both for $J{=}0$ and $J{=}1$. Nuclear spin-rotation interactions as well as the direct dipole-dipole interaction between the nuclear spins  \cite{Brown:2003} were found to give negligible contributions. Fitting this model to the observed spectra yields the molecular constants summarized in Table~\ref{table:const}.

%% E FIELD 

When applying electric fields, already above few tens of V/cm, the complexity of the rotational spectra reduces significantly, as shown in Fig.~\ref{fig:Bfield}(b). In this regime, the Stark effect $H_{\rm S} = - \vec{d} \cdot \vec{E}$ dominates over the hyperfine interactions, where $\vec{d}$ denotes the permanent electric dipole moment of the molecules and $\vec{E}$ the electric field. The $m_J{=}0$ state separates from the now degenerate pair $m_J{=}\pm1$, while the nuclear spin projections $m_{I_{\rm Na}}$ and $m_{I_{\rm K}}$ decouple from $m_J$. Accordingly, only the three transitions with $\Delta m_J = 0,\pm 1$ that do not change $m_{I_{\rm Na}}$ and $m_{I_{\rm K}}$ are accessible.

%%%%%%% FIGURE 3 - J=1 lifetime %%%%%%%

\begin{figure}
  \begin{center}
  \includegraphics[width=1\columnwidth]{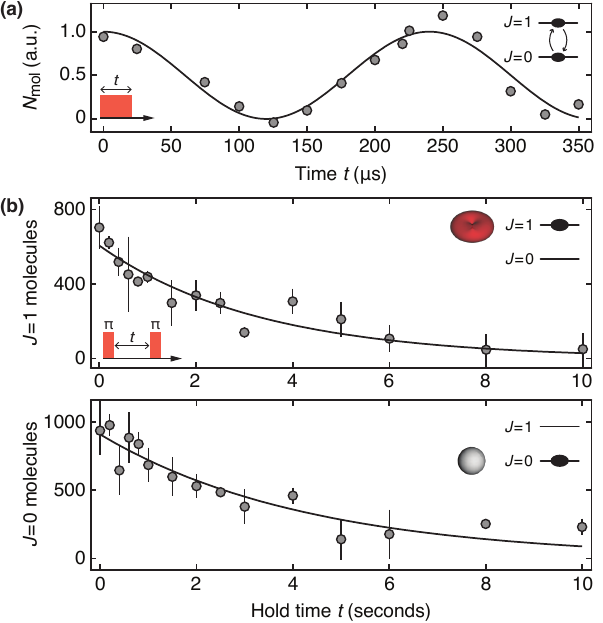}
  \end{center}
  \caption{\label{fig:lifetime} (color online) Coherent transfer and collisional lifetime of chemically stable $^{23}$Na$^{40}$K molecules in the $J{=}1$ state. (a) High-contrast Rabi oscillations between $|0,0, 3/2,-4\rangle$ and the lowest energy hyperfine state of $J{=}1$ at a magnetic field of $85.6\, \rm G$. The $\pi$-pulse duration is $120\, \mu \rm s$. (b) Lifetime of $^{23}$Na$^{40}$K in the $J{=}1$ ($J{=}0$) state, shown in the upper (lower) panel. Solid lines show exponential fits $ A \, e^{-t/\tau}$, and yield comparable lifetimes $\tau$ of $3.3(4)\, \rm s$ and $4.6(7)\, \rm s$, respectively. Data points show the average of typically three experimental runs; the error bars denote the standard deviation of the mean. %\red{1 Lifetime has exponential fit for now. 2 Give K3 coefficient, density.}  
}
\end{figure}

%%%%%%%%%%%%%%%%%%%%%%%%%%%%%%%%

Equipped with the understanding of rotational transitions, we can coherently manipulate the internal quantum states of the trapped $^{23}$Na$^{40}$K molecules. As a first application, we demonstrate coherent population transfer between $J{=}0$ and $J{=}1$, as shown in Fig.~\ref{fig:lifetime}. Using a microwave $\pi$-pulse to transfer the entire molecular ensemble into the lowest hyperfine state of $J{=}1$, we observe a remarkable lifetime of $3.3(4)\, \rm s$, which is comparable to $4.6(7)\, \rm s$ measured in the $J{=}0$ state; both data sets were taken for an initial peak density of $0.7 \times 10^{11}\, \rm cm^{-3}$ and a temperature of $400\, \rm nK$. This demonstrates that even in a rotationally excited state, dense ensembles of ultracold molecules can be collisionally stable. We point out that the van der Waals interactions between molecules in $J{=}1$ are significantly less attractive than in $J{=}0$, as the corresponding $C_6$ coefficients are dominated by virtual transitions to nearby rotational states \cite{Lepers2013}. Since the eigenenergies of rotational states can be controlled by the Stark effect [see Fig.~\ref{fig:Bfield}(b)], van der Waals interactions in $J{=}1$ can be tuned from attractive to repulsive via external electric fields \cite{avdee2006}. 

%%%%%%% FIGURE 4 - Two MW pulse spectroscopy %%%%%%%

\begin{figure}
  \begin{center}
  \includegraphics[width=1\columnwidth]{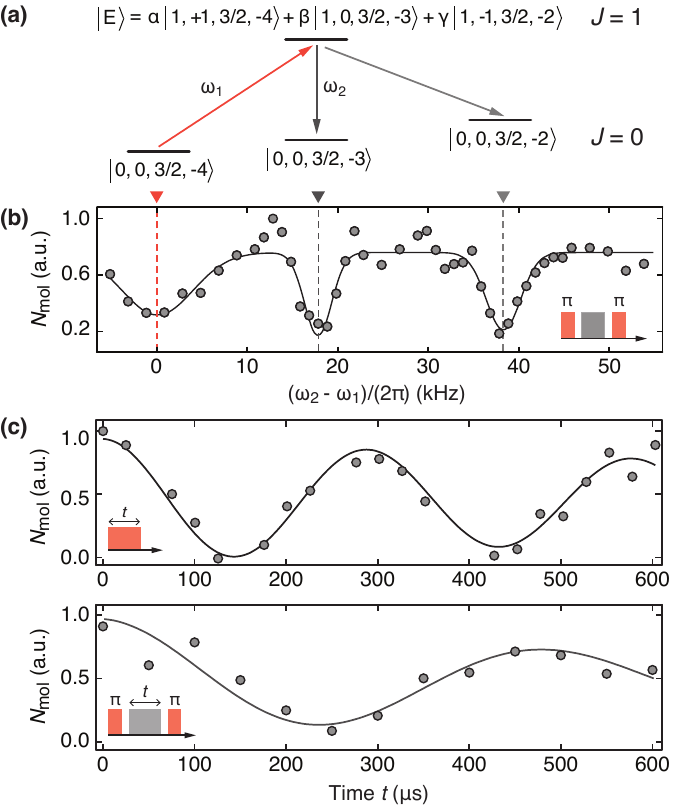}
  \end{center}
\caption{\label{fig:twopulse} Coherent population transfer between hyperfine states of the rotational ground state $J{=}0$. (a) A first $\pi$-pulse at frequency $\omega_{1}$ transfers the population from the hyperfine ground state $|0,0, 3/2, -4 \rangle$ to a rotationally excited state $|\rm{E}\rangle$ of mixed $m_{I_{\rm K}} = -4,\,-3,\,-2$ character, but fixed $m_{I_{\rm Na}} = 3/2$. A second microwave pulse at frequency $\omega_{2}$ coherently couples $|\rm{E}\rangle$ to the hyperfine states $|0,0, 3/2, -2 \rangle$ or $|0,0, 3/2, -3\rangle$. (b) Hyperfine spectrum of the rotational ground state at $85.6\, \rm G$. Following the initial $\pi$-pulse at $\omega_{1}$, the population of $|\rm{E}\rangle$ is monitored as a function of $\omega_2$. Dashed vertical lines indicate the observed resonances. The solid line serves as a guide to the eye. The strength of the $|0,0,3/2,-2\rangle$ resonance is enhanced by the presence of the optical dipole trap ($\lambda = 1064\, \rm nm$). (c) Rabi oscillations between $|0,0, 3/2, -4 \rangle$ and $|\rm{E}\rangle$ (top), and $|\rm{E}\rangle$ and $| 0,0, 3/2, -3 \rangle$ (bottom). The microwave power was identical for both data sets.
}
\end{figure}

%%%%%%%%%%%%%%%%%%%%%%%%%%%%%%%%

%% STRONG ROTATIONAL TRANSITIONS (CHECK WHETHER THIS IS CORRECT)

In the singlet rovibrational ground state of $^{23}$Na$^{40}$K, the hyperfine states are highly attractive for the storage of quantum information. Due to the absence of electron spin, only nuclear spins remain and give rise to comparably small magnetic moments. This makes superpositions of such hyperfine states inherently insensitive to magnetic field noise. However, the same small magnetic moments hamper the creation of hyperfine superpositions in $J{=}0$ via direct magnetic spin-flip transitions. In contrast, transitions between rotational states involve the large electric dipole moment of the NaK molecule, and Rabi frequencies for a given electromagnetic wave are about $ m_{\rm p}/( m_{\rm e} \alpha) \sim 10^5$ times higher than magnetic dipole transitions. Therefore, two consecutive rotational transitions $J{=}0 \rightarrow J{=}1 \rightarrow J{=}0$ or coherent two-photon transitions can efficiently create superpositions of hyperfine states in $J{=}0$~\cite{Ospelkaus2010}.

%% COHERENT COUPLING VIA J=1

Figure~\ref{fig:twopulse} demonstrates that mixing of nuclear spins within the first rotationally excited state can serve as a bridge to coherently manipulate hyperfine states within the rotational ground state of $^{23}$Na$^{40}$K. While $J{=}0$ states above $2\,\rm G$ have defined nuclear spin projections $ m_{I_{ \rm Na}}$ and $m_{I_{\rm K}}$, $J{=}1$ states have mixed spin character as  nuclear quadrupole coupling remains significant for magnetic fields up to several hundred G. Using two consecutive microwave pulses, the initial hyperfine state $|0,0,3/2, -4\rangle$ can first be transferred to an intermediate mixed state $|{\rm E}\rangle$, and subsequently coupled to a different hyperfine state of $J{=}0$. The spectrum in Fig.~\ref{fig:twopulse}(b) shows that the mixing in $J{=}1$ can be sufficiently strong to access several hyperfine states in the rotational ground state. We compare the coupling strengths of the first and the second transition by recording Rabi oscillations [see Fig.~\ref{fig:twopulse}(c)]. For identical microwave powers, the Rabi frequencies $\Omega_1 \approx 2 \pi \times 3.4\, \rm kHz$ and $\Omega_2 \approx 2 \pi \times 2.1\, \rm kHz$ are comparable, indicating similar magnitudes for the amplitudes $\alpha$ and $\beta$. This opens the possibility of driving efficient, direct two-photon transitions between two $J{=}0$ hyperfine states.

%%%%%% CONCLUSION

In conclusion, we have demonstrated coherent microwave control of rotational and hyperfine states of ultracold $^{23}$Na$^{40}$K molecules. In particular, we have transferred the entire molecular sample to the first rotationally excited state and revealed that collisional lifetimes in $J{=}1$ can be long,  comparable to those in the $J{=}0$ state. Utilizing the strongly mixed nuclear spin character in the $J{=}1$ state, we have coherently transferred population between hyperfine states of the rotational ground state. Achieving full control over internal degrees of freedom is a crucial step towards applications of dipolar molecules for the realization of novel many-body phenomena, such as SU(N) symmetric physics and topological superfluidity. Microwave and electric field control of long-lived rotational states allows the engineering of collisional properties of dipolar molecules, which may facilitate the evaporative cooling of chemically stable molecules to reach quantum degeneracy.

We would like to thank David DeMille, Robert Field and collaborators, and Jeremy Hutson for fruitful discussions. This work was supported by the NSF, AFOSR PECASE, ARO, an ARO MURI on ``High-Resolution Quantum Control of Chemical Reactions'', an AFOSR MURI on ``Exotic Phases of Matter'', and the David and Lucile Packard Foundation. Z.Z.Y.~acknowledges additional support by the NSF GRFP.

\bibliographystyle{apsrev4-1}
\bibliography{NaK_ref_v2}

\end{document}